\documentclass[traditabstract]{aa}
\usepackage{txfonts}
\usepackage{graphicx}
\usepackage{natbib}
\usepackage[dvips]{color}

\bibpunct{(}{)}{;}{a}{}{,}

\begin{document}

\newcommand{\be}{\begin{equation}}
\newcommand{\ee}{\end{equation}}
\newcommand{\bdm}{\begin{displaymath}}
\newcommand{\edm}{\end{displaymath}}
\newcommand{\bea}{\begin{eqnarray}}
\newcommand{\eea}{\end{eqnarray}}

\newcommand{\cf}{\textit{cf.}~}
\newcommand{\ie}{\textit{i.e.}~}

\newcommand{\oz}[1]{ \textcolor{red}   {\texttt{\textbf{OZ: #1}}} }

\title{Model for an optically thick torus in local thermodynamic equilibrium
around a black hole}

\author{
        O. Zanotti    \inst{1}
}

\institute{
Universit\`a di Trento, Laboratorio di Matematica Applicata, Via Messiano 77, I-38123 Trento, Italy
\\ \email{olindo.zanotti@unitn.it}
}

\date{}

\authorrunning{O.~Zanotti}
\titlerunning{
Optically thick torus around a black hole.}

\abstract
{
We propose a simple model for an optically thick radiative
torus in local thermodynamic  equilibrium
around a Kerr black hole. The hydrodynamics
structure, which is not affected by the radiation field,
is the same as for 
the so--called {\em polish doughnuts}.
Under the assumption of isentropic fluid and polytropic
equation of state, 
a simple stationary and axisymmetric solution to the relativistic
radiation hydrodynamics equations is possible, for which
the  temperature of the torus 
scales like the specific enthalpy. The astrophysical relevance of
the model is briefly discussed.
}


\maketitle


\section{Introduction}
\label{sec:Introduction}
Relativistic radiation hydrodynamics provides the natural framework
of several high-energy astrophysical processes, essentially all those for
which the interaction of photons with matter takes place
in a strong gravity regime or in the presence of
relativistic motions. One such relativistic system is
represented by geometrically thick disks (tori) around
black holes. 
Considered as inviscid purely fluid solutions, 
these objects have always attracted much interest,
partly because they allow for relatively simple analytic
or semi--analytic
configurations~\citep{Fishbone76,Abramowicz78,Kozlowski1978,Font02a,Daigne04,Qian2009,Penna2013b}
and partly because they can be adopted to study
various types of fluid instabilities and potentially
observable physical effects
in the vicinity of black holes~\citep{Abramowicz1980,Abramowicz83,Papaloizou84,Abramowicz1998,
  Rezzolla_qpo_03b, Zanotti03, Blaes2006, Montero2010}. 
Moreover, a renewed interest for them has been motivated by the
outcome of recent numerical simulations in full general
relativity, showing that
high-density tori are indeed produced after the
merger of unequal--mass neutron star binaries that form a
black hole~\citep{Rezzolla:2010}.

In spite of all these analyses, and although tori
have been extensively studied
in numerical simulations
also for the effects of
magnetic
fields~\citep{DeVilliers03,Komissarov2006a,Fragile2009b,
Barkov2011,Narayan2012,
McKinney2012}, the role of
radiation fields has often been disregarded.
This has certainly been due to the 
complexity inherent in the time--dependent
solution of the relativistic
radiation hydrodynamics equations, which, at a rigorous
level, would require the solution of the 
Boltzmann equation for the
distribution function of photons.\footnote{See 
\citet{McClarren10} and \citet{Radice2013} for some
promising progress in this direction.} 
For this reason,
a few more pragmatic approaches have been developed over the
years,  which become useful both from a theoretical and
from a numerical point of view. 
Perhaps the most successful one is represented by 
the so--called projected symmetric
trace--free (PSTF) moment formalism of
\cite{Thorne1981}, which defines
the moments of the
radiation field similarly to how density, momentum and pressure of a
fluid are defined as moments of the
corresponding distribution function. 

In the recent past, by adopting this procedure, 
a number of time-dependent numerical codes have been
developed and subsequently applied to 
various scenarios related to accretion
flows in a relativistic
context~\citep{Shapiro96,Farris08,Zanotti2011,Fragile2012,
Roedig2012,Sadowski2013,Takahashi2013}. 
\citet{Sadowski2013b}, in particular, have studied the
super--critical accretion onto a black hole from a
radiative torus, showing that most of the luminosity
emerges through the funnels and confirmed, broadly
speaking, the physical importance of these objects for
high-energy astrophysics after so many years of active
research. 

The goal of this paper is to provide a simple 
model for an optically thick radiative
torus around a black hole. Local thermodynamic equilibrium is assumed,
and it is shown that a radiation field can be added while
keeping the underlying hydrodynamic solution unmodified.
The new model can be useful both on theoretical grounds,
for clarifying the physics of radiative tori, and
on numerical grounds, since it offers a test for general
relativistic radiation codes in the optically thick
regime. However, the time--dependent numerical
investigation and the study of the stability properties
of the proposed model are not the focus of this work and will
be considered elsewhere.

In the following, we set the speed of light $c=1$,
and the gravitational constant $G=1$.
We extend the geometric units by setting $m_p/k_B=1$,
where $m_p$ is the mass of the proton, while $k_B$ is the Boltzmann
constant. 
However, we have maintained $c$ in an explicit form in expressions of particular physical interest.

\section{Physics of the model}
\label{sec:Physics-of-the-model}

\subsection{Basics assumptions}
We considered a stationary and axisymmetric space-time
around a Kerr black hole, which may be written either in
Boyer-Lindquist or in Kerr-Schild coordinates.
%
%
The energy--momentum tensor of the system is given by
\be
T^{\alpha\beta}=T^{\alpha\beta}_{{\rm m}}+T^{\alpha\beta}_{{\rm r}}\,,
\ee
which clearly contains two contribution. The first term
is due to matter and can be written as
\be
T^{\alpha\beta}_{{\rm m}}=\rho
h\,u^{\,\alpha}u^{\beta}+pg^{\,\alpha\beta}\,,
\label{T_matter}
\ee
where $g^{\alpha\beta}$ is the metric of the background space-time, $u^\alpha$ is
the four--vector velocity of the fluid, while $\rho$, $h=1+\epsilon + p/\rho$,
$\epsilon$ and $p$ are the rest--mass density, the specific (i.e. per
unit mass) enthalpy,
the specific internal energy, and the pressure,
respectively. For the time being, we keep the equation of
state unspecified. The fluid is in circular motion around
the black hole and its four--vector velocity is
$u^\alpha=u^t(1,0,0,\Omega)$, where $\Omega=u^\phi/u^t$ is
the angular velocity as observed from infinity. 

The second contribution to the energy--momentum tensor is
due to the radiation field, and, in terms of its moments,
can be written as \citep{Hsieh1976}
\be
\label{T_rad1}
T_{\rm r}^{\alpha\beta}=(E_{\rm r}+{\cal P}_{\rm r}) u^\alpha u^\beta +
F_{\rm r}^\alpha u^\beta+ u^\alpha F_{\rm r}^\beta + P^{\alpha\beta}_{\rm r}\,,
\ee
where the energy density, the radiation flux and the radiation
stress tensor, all measured in the comoving frame of the fluid, are given by
\bea
&&E_{\rm r}=\frac{1}{c}\int I_\nu d\nu d\Omega \,, \\
&&F_{\rm r}^\alpha=h^\alpha_{\ \beta}\int I_\nu d\nu d\Omega N^\beta
\,, \\
&&P^{\alpha\beta}_{\rm r}=\frac{1}{c}\int I_\nu d\nu d\Omega N^\alpha N^\beta
\,.
\eea
We recall that
$I_\nu=I_\nu(x^\alpha,N^i,\nu)$ is the specific (i.e. an
energy flux per unit time, frequency, and solid angle)
intensity of the radiation, 
 $N^\alpha$ is the four--vector defining the photon
propagation direction, 
$d\nu$ is the infinitesimal frequency and
$d\Omega$ is the infinitesimal solid angle around the direction of
propagation. Finally,
$h^{\alpha\beta}=u^\alpha u^\beta + g^{\alpha\beta}$ 
is the projector operator into
the space orthogonal to the four--velocity $u^\alpha$.
In full generality, the four--force density of the radiation field, which
describes the interaction of radiation with matter, 
is given by \citep{Mihalas84,Shapiro96}
\be
G^\alpha_{\rm r}=\frac{1}{c}\int(\chi_\nu I_\nu-\eta_\nu)N^\alpha
d\nu d\Omega \,,
\ee
where $\chi_\nu \equiv \chi_\nu^t+\chi_\nu^s$ and $\eta_\nu\equiv
\eta_\nu^t+\eta_\nu^s$ are the total opacity and emissivity
coefficients,
each containing a thermal contribution, indicated with superscript
t, and a scattering contribution, indicated with superscript
s. However, if additionally assume that
the fluid is optically thick and in local
  thermodynamic equilibrium with the radiation,
then the radiation four--force vanishes, and the energy
density is just $E_{\rm r}=4\pi \tilde{B}=a_{\rm
  rad}T^4$,
where $4\pi \tilde{B}=a_{\rm rad}T^4$ is the equilibrium black--body
intensity, $T$ is the local temperature of the fluid, and $a_{\rm rad}$ is
the radiation constant.
Under these conditions, the radiation is isotropic in the
comoving frame of the fluid, the radiation pressure is 
${\cal P}_{\rm r}=E_{\rm
  r}/3,$ and the radiation fluxes $F_{\rm r}^\alpha$ also
vanish~\citep{Farris08}.
As a result, the energy--momentum tensor
(\ref{T_rad1}) of
the radiation field reduces to 
\be
\label{T_rad2}
T_{\rm r}^{\alpha\beta}=\frac{4}{3}E_{\rm r}u^\alpha u^\beta +
\frac{1}{3}E_{\rm r}g^{\alpha\beta}\,.
\ee

\subsection{Description of the model}
We searched for a stationary, that is, $\partial_t\equiv0$,  and
axisymmetric,  that is, $\partial_\phi\equiv0$,
solution to the
full set of equations describing the dynamics of the
system, that is,
\bea
\label{0eq:mass}
&&\nabla_{\alpha} (\rho u^{\,\alpha})=0, \\
\label{0eq:momentum}
&&\nabla_{\alpha}T^{\alpha\beta}=0, \\
\label{0eq:rad}
&&\nabla_\alpha T_{\rm r}^{\alpha\beta}=-G^\beta_{\rm r} \,.
\eea
The continuity equation is of course trivially satisfied
when the velocity field is purely toroidal. 
In addition, since local thermodynamic equilibrium is assumed,
and the radiation four--force vanishes,
Eq.~(\ref{0eq:momentum}) effectively decouples into 
\be
\label{1eq:momentum-a}
\nabla_{\alpha}T_{\rm m}^{\alpha\beta}=0
\ee
and
\be
\label{1eq:momentum-b}
\nabla_{\alpha}T_{\rm r}^{\alpha\beta}=0\,,
\ee
which must be satisfied separately.
From (\ref{1eq:momentum-a}), after contracting with the operator
$h^{\alpha\beta}$, we can write the Euler equation in the
form~\citep{Abramowicz78}
\begin{equation}
\label{accel:gas}
u^\alpha\nabla_\alpha u_\beta = -\frac{\nabla_\beta
  p}{h\rho}=\nabla_\beta \ln |u_t| -
\left(\frac{\Omega}{1-\Omega\ell}\right)
\nabla_\beta\ell
\,,
\end{equation}
where the last term is the four--acceleration of the
fluid in circular motion.
This equation was extensively studied in the context of
geometrically thick disks around black holes, and
depending on the specific\footnote{When referred to the
  angular momentum, the adjective {\em specific} means
 ``per unit energy''.} angular momentum
$\ell=-u_\phi/u_t$, different classes of equilibrium
tori can be obtained\footnote{
See \citet{Rezzolla_book:2013} for a pedagogic exposition.}
whose structure and
properties have been discussed in detail
\citep{Abramowicz78,Kozlowski1978,Font02a,Daigne04}. 
Here we investigate the possibility of adding an
optically thick radiation field to those solutions
without affecting their hydrodynamics.
This extension is possible, but
only for a limited class of equations of state. First of
all, we contract Eq.~(\ref{1eq:momentum-b})
with  $h^{\alpha\beta}$ to find, after a few
tensor operations, that 
\begin{equation}
\label{accel:rad}
u^\alpha\nabla_\alpha u_\beta = -\frac{\nabla_\beta
  E_{\rm r}}{4E_{\rm r}}\,. 
\end{equation}
To make (\ref{accel:gas}) and (\ref{accel:rad})
compatible, it must be
\begin{equation}
\label{eq:p-E}
\frac{\nabla_i p}{h\rho}=\frac{\nabla_i
  E_{\rm r}}{4E_{\rm r}}\hspace{1cm}{\rm for}\hspace{0.2cm} i=r,\theta\,.
\end{equation}
We now need to address the thermodynamic
properties of the fluid. We limited our
attention to isentropic fluids, for which $dp=\rho dh$.
This has two important consequences. The first one
is that the specific enthalpy can be computed as
\begin{equation}
\label{eq:h_vs_W}
h = \exp({\mathcal{W}_\mathrm{in}-\mathcal{W}})\,,
\end{equation}
which follows from the second equality of Eq.~(\ref{accel:gas})
after introducing the potential $\mathcal{W}$ as \citep{Kozlowski1978}
\begin{equation}
\label{eq:equilibrium0}
\mathcal{W}-\mathcal{W}_\mathrm{in} = \ln{|u_{t}|}-\ln{|(u_{t})_\mathrm{in}|}-
\int_{\ell_\mathrm{in}}^{\ell}\frac{\Omega\, d\ell'}{1-\Omega \ell'}\,.
\end{equation}
\begin{table*}
\begin{center}
\caption{Main parameters of 
a few representative  tori in
local thermodynamic equilibrium with a radiation field.
>From left to
right the columns report the name of the model, 
the black hole spin $a$,
 the power--law index $q$ of the angular momentum
distribution (when pertinent),
the constant specific
angular momentum $\ell$ (when pertinent), the inner and outer radii 
$r_\mathrm{in}$ and
$r_\mathrm{out}$,  
the radial position of the
center $r_\mathrm{c}$,
and the orbital period at the center of the torus
$t_\mathrm{orb}$,
The last column reports the temperature ratio
$T_{\rm c}/T_{\rm in}$.
All models share
the same mass for the black hole, $M=2.5M_{\odot}$,
polytropic exponent 
$\Gamma=4/3,$ and disk-to-hole mass
ratio $M_{\rm t}/M=0.1$.
}
\label{tab1}
\begin{tabular}{l|lcccccccc}
\hline
Model   & $a$           &
 $q$& $\ell$  & $r_{\rm in}$     & $r_{\rm out}$ 
        & $r_{\rm c}$ & $t_{\rm orb}$  & $T_{\rm c}/T_{\rm in}$
          \\
&  & & & & & & ${\rm (ms)}$ 
          & &\\

\hline
(a)   & 0.0     & $-$ & 3.7845  &
        4.646 & 14.367& 8.165 & 1.80    & 1.0079\\
(b)   & 0.0      & $-$ & 3.8022  &
        4.566 & 16.122 & 8.378 & 1.87   &  1.0099\\
(c)   & 0.0      & $-$ & 3.8425 
        &4.410 & 21.472 & 8.839 &2.03    &  1.0152\\
(d)   & 0.0      & $-$ & 3.8800 
        &4.290 &29.539& 9.246 &2.17   &   1.0209\\
\hline
(e)   & 0.7     & 0.1 & $-$ 
        &3.004 & 11.032 & 5.556 & 1.06  &   1.0132 \\
(f)   & 0.9     & 0.1 & $-$ 
        &1.971 & 16.535 & 4.124 & 0.71  &   1.0485\\
\end{tabular}
\end{center}
\end{table*}
Note that subscript $\mathrm{in}$ refers to the inner edge of the
disk in the equatorial plane, where $h_\mathrm{in}=1$.
The second consequence of the isentropic assumption
is that
Eq.~(\ref{eq:p-E}) can be integrated to give
\begin{equation}
\label{eq:h-E}
h=\left(\frac{E_{\rm r}}{E_{\rm r,in}}\right)^{1/4}=\frac{T}{T_{\rm in}}\,,
\end{equation}
where $E_{\rm r,in}=a_{\rm rad} T_{\rm in}^4$ is the energy density of the
radiation field at $r=r_{in}$, while $T_{\rm in}$ is the
surface temperature.
Eq.~(\ref{eq:h-E}) poses some restrictions on the
equations of state that can be accepted. For instance, an
ideal fluid equation of state $p=\rho\epsilon(\gamma-1)$, 
for which $T=p/\rho$ and $h=1+\gamma/(\gamma-1)T$, would imply 
a constant temperature from Eq.~(\ref{eq:h-E}). This is
clearly not
acceptable, since $p/\rho$ is not uniform within the torus.
Hence, we need an equation of state for which the
temperature is a free parameter.
This is the case of the
polytropic equation of state
\begin{equation}
\label{polytrope-Gamma}
p=K\rho^\Gamma\,,
\end{equation}
where $K$ is the polytropic constant and $\Gamma$ is the
adiabatic index of the polytrope.
We stress
that the equation of state of  isentropic  polytropes
can still be written formally like
$p=\rho\epsilon(\Gamma-1)$. However, there are two
important differences with respect to 
the equation
of state of an ideal fluid, even when the latter is isentropic.
The first difference is that
the exponent $\Gamma$ of Eq.~(\ref{polytrope-Gamma}) is not
necessarily 
given by the ratio of the specific heats at constant
pressure and at constant volume. The second difference is
that for polytropes like (\ref{polytrope-Gamma}),
the temperature $T$ is not
necessarily given by the ratio $p/\rho$ and can therefore
be regarded as a free parameter.\footnote{A famous example is
given by the degenerate Fermi
fluid, whose equation of state  can be written like
Eq.~(\ref{polytrope-Gamma}).
}
It is interesting to note that while a purely
hydrodynamic torus is compatible with the ideal fluid
equation of state, in which case its temperature is given by $p/\rho$,
the radiation hydrodynamic torus that we are describing
is not.

After adopting the equation of state (\ref{polytrope-Gamma}),
Eq.~(\ref{eq:h-E}) can be used to fix the temperature of
the torus as
\begin{equation}
\label{eq:T-h}
T=T_{\rm in} h\,.
\end{equation}
Hence, 
the isocurves of the temperature
$T$ and those of the specific enthalpy $h$ of the torus
coincide.

A well-known property of polish doughnuts is that they
possess
an internal Keplerian point, $r_{\rm c}$,
lying on the equatorial
plane, where the rest--mass density reaches a maximum
$\rho_{\rm c}$. This also corresponds to the highest
temperature of the torus, and the ratio of the central to
the surface temperature is given by
\begin{equation}
\label{eq:T-ratio}
\frac{T_{\rm c}}{T_{\rm in}}=h_{\rm c}=
1+\frac{\Gamma}{\Gamma-1}\frac{p_{\rm c}}{\rho_{\rm c}}=
\exp({\mathcal{W}_\mathrm{in}-\mathcal{W}_\mathrm{c}})>1\,.
\end{equation}

We note that the surface
temperature $T_{\rm in}$, or, alternatively, the central
temperature $T_{\rm c}$, is the only missing
parameter that must be specified in addition to those
necessary for constructing the hydrodynamic model.
We also stress that since the temperature scales like
the specific enthalpy, and the specific enthalpy is given
by Eq.~(\ref{eq:h_vs_W}), the temperature does not
depend on the polytropic constant $K$ or on the index
of the polytrope $\Gamma$. This is reminiscent of a
similar result originally proved by \citet{Rezzolla_qpo_03b} for
purely hydrodynamic relativistic tori,
namely that they have ratios $p/\rho,$ which do not depend on the
polytropic constant $K$.

Finally, a consistency check can be performed to show that even
in the presence of a temperature gradient, the
relativistic radiation flux
is zero. In fact, 
in the framework of Eckart's formulation of relativistic
standard irreversible thermodynamics~\citep{Eckart40c}, 
the  flux is given by the relativistic form of Fourier's law, namely
\citep{Israel76}
\be
\label{heat-flow}
F_{\rm r}^\mu=-\lambda T(h^{\mu\nu} \nabla_\nu\ln T + a^\mu)\,,
\ee
where 
$\lambda$ is the thermal conductivity.
It is easy to see that if the four--vector acceleration is given
by the right-hand side of Eq.~(\ref{accel:rad}), $F_{\rm rad}^\mu$
as given by Eq.~(\ref{heat-flow}) is indeed zero.

\subsection{A few representative examples}
As a first illustrative example, we consider a
Schwarzschild space-time
with the fluid having a constant (in radius) specific
angular momentum $\ell$.
In these circumstances the rest--mass density distribution can be solved
analytically and is~\citep{Font02a}
\begin{equation}
 \label{eq:rho_vs_W}
\rho  = \left[\left(\frac{\Gamma-1}{K \Gamma}\right)
\left[\exp(\mathcal{W}_\mathrm{in}-\mathcal{W})
-1\right]\right]^{1/(\Gamma-1)}\,,
\end{equation}
where
\begin{equation}
\label{eq:Weq_a=0}
\mathcal{W}(r,\theta) = \ln|u_t| = \frac{1}{2} \ln \left[
\frac{r^2(r-2)\sin^2\theta}{r^3\sin^2\theta - \ell^2(r-2)} \right]
\end{equation}
is the potential.
If we  fix the position of the inner edge of
torus $r_{\rm in}$ at the position of the cusp $r_{\rm
  cusp}$ of the potential $\mathcal{W}(r,\pi/2)$,
the torus will exactly fill its Roche lobe.
Under this assumption, we have considered a few 
tori orbiting around a
$M=2.5M_{\odot}$ black hole.
The first four rows of Table~\ref{tab1} report 
four models previously considered by \citet{Zanotti03} in
the Schwarzschild space-time, but without the
radiation field.\footnote{
Models (a)--(d) in Table~\ref{tab1}
are the same as models (e)--(h) of \citet{Zanotti03}.
}
All the models have the same disk-to-hole
mass ratio $M_t/M=0.1$, but different sizes.
The last column
reports the ratio $T_{\rm c}/T_{\rm in}$, computed
according to Eq.~(\ref{eq:T-ratio}).
As can be seen, tori that are in local thermodynamic equilibrium
and obey Eq.~(\ref{eq:h-E}) are close to be
isothermal, with deviations from $T={\rm const}$ that
increase as the inner edge of the torus approaches the
black hole.

The property of being almost isothermal remains true also
when power-law distributions of the specific
angular momentum are adopted, namely when
$\ell (r,\theta = \pi/2) = {\cal S} r^q$, although 
a richer phenomenology is possible for them.\footnote{
See \citet{Daigne04} for an extended discussion about
tori with a power-law distribution of the specific
angular momentum.}
We recall that in the Kerr metric the radius $r_{\rm  ms}$ of the marginally stable orbit
is a decreasing function of the black-hole spin parameter
$a$. On the other hand, when the 
power law-index $q$ is increased, the inner radius of the torus increases as well. Hence, 
for low values of $q$, the torus 
penetrates deeper into the potential well, placing the cusp below the marginally
stable orbit. In these circumstances the highest
values of $T_{\rm c}/T_{\rm in}$ are found.
The last two rows of Table~\ref{tab1} report 
two representative models of
this class,
which were previously considered by \citet{Zanotti05} in Kerr space-time, but without the
radiation field.\footnote{
Models (e)--(f) in Tab.~\ref{tab1}
are the same as models (D2a) and (E1a) of \citet{Zanotti05}.
}
  
In contrast, when
the exponent $q$  is close to $1/2$,
the rotation law tends to the Keplerian one, with a
torus that flattens
toward the equatorial plane. 
Tori of this kind are generally much larger
than $\ell ={\rm const}$ tori \citep{Zanotti2010},
with radii at the maximum rest--mass density 
point as high as $r_{\rm c}\sim1000$. Consequently,
they are very mildly
relativistic, with values of the specific enthalpy 
$h_{\rm c}$ very close to unity. We have
verified this trend for a few representative cases, not
reported in Table~\ref{tab1}, and found that
$T_{\rm c}/T_{\rm in}\approx 1$ to a
few parts over $10^6$.

\section{Discussion and conclusions}
\label{sec:conclusions}
We have shown that purely hydrodynamical geometrically
thick disks around a black hole, 
extensively studied in the past and sometimes referred to
as {\em polish doughnuts},
can be {\em dressed} with a
radiation field in local thermodynamic equilibrium with the fluid.
The equation of
state required for this extension is that of an
isentropic and polytropic fluid, with a temperature that
scales like the specific enthalpy. The resulting models,
computed for a few representative cases of both constant
and varying distributions of the specific angular
momentum, are almost isothermal, with
deviations of only a few percent.
 
Since the equation of state of the model is nonthermal
and well--suited to describe degenerate matter, 
the most natural astrophysical scenario 
where such a model can become relevant
is that of a high--density
torus, with densities close to those of neutron stars.
Although still rather simplified,
these models may indeed offer a better description 
of high--density tori produced after the merger of
unequal--mass neutron star
binaries collapsing onto a black hole, as reported by
fully relativistic and purely hydrodynamical
numerical simulations~\citep{Rezzolla:2010}.
This marks an important difference with respect to the
radiation torus discussed in \citet{McKinney2013b}, which
adopts a thermal equation of state and is instead well--suited to
describe (much) lower density accretion disks around
black holes.

The model discussed here may also offer
a stationary test for
numerical  codes that solve the equations of general
relativistic radiation hydrodynamics.
However, a few words of caution need to be said 
in this respect.
The first warning
is that since the torus is isentropic,
the evolution should preferably be performed 
without integrating the energy density of
the fluid, which can be computed algebraically
from the remaining quantities by exploiting
Eq.~(\ref{polytrope-Gamma}). Secondly, since the opacity
is typically a function of the rest--mass density, which drops at the surface of the torus,
the
assumption of rigorous optical thickness is likely 
to fail in a thin shell at the surface of the
torus. However, this problem is mitigated by the
possibility of building models for which such a shell
is arbitrarily thin. Thirdly, any modern numerical code for
the solution of the general
relativistic radiation hydrodynamics equations will require
the treatment of a tenuous ``atmosphere'', that is, of a
low--density region filling the supposedly vacuum space
around the torus. The atmosphere is
intrinsically optically thin and this requires a code
capable of handling both the optically thick and the optically thin
regimes. Significant progress in this respect has recently been
obtained by \citet{Sadowski2013,Takahashi2013,McKinney2013b}.

Finally, the stability properties of these new models
and their response to perturbations remain an open question,
which may deserve a dedicated
investigation, both analytical and numerical.

\begin{acknowledgements}
I thank John C.~Miller and the
referee Jonathan McKinney for very useful suggestions that
helped me to improve the quality of the manuscript.
This work  has been financed in parts by the European Research Council under 
the European Union's Seventh Framework Programme (FP7/2007-2013) in the frame of the research 
project \textit{STiMulUs}, ERC Grant agreement no. 278267.
\end{acknowledgements}

\bibliographystyle{aa}
\bibliography{biblio/aeireferences}

\end{document}